\documentclass{wqcd03}                 
\newcommand{\tev}{\mbox{TeV}}
\newcommand{\gev}{\mbox{GeV}}

\def    \be             {\begin{equation}}
\def    \ee             {\end{equation}}
\def    \ba             {\begin{eqnarray}}
\def    \ea             {\end{eqnarray}}
\def    \nn             {\nonumber}
\def    \=              {\;=\;}
\def    \frac           #1#2{{#1 \over #2}}

\def    \ie             {{\em i.e.\/} }
\def    \eg             {{\em e.g.\/} }

\def    \pt             {\mbox{$p_T$}}
\def    \et             {\mbox{$E_T$}}

\def    \qsq           {\mbox{$Q^2$}}

\newcommand     \ptmin     {\ifmmode p_{\scriptscriptstyle T}^{\sss min} 
                           \else    $p_{\scriptscriptstyle T}^{\sss min}$ \fi}

\newcommand     \MSB            {\ifmmode {\overline{\rm MS}} \else 
                                 $\overline{\rm MS}$  \fi}

\def    \mur            {{\mbox{$\mu_{\rm R}$}}}

\def    \as             {\ifmmode \alpha_s \else $\alpha_s$ \fi}
\def    \asmz             {\ifmmode \alpha_s(M_Z) \else $\alpha_s(M_Z)$ \fi}

\def    \Qbar   {\overline{Q}}
\def    \qbar   {\overline{q}}

\newcommand \cpviol{\ifmmode{{\rm CP}\!\!\!\!\!\!\! \not\phantom{P}}\else{${\rm
      CP}\!\!\!\!\!\!\! \not \phantom{P}$}\fi}
\def\met{$\rlap{\kern.2em/}E_T$}
\def\herwig{{\small HERWIG}}
\def\isajet{{\small ISAJET}}
\def\pythia{{\small PYTHIA}}
\def\ariadne{{\small ARIADNE}}

\def\MCNLO{{\small MC@NLO}}
\def\sherpa{{\small SHERPA}}

\confname{QCD@Work 2003 - International Workshop on QCD, Conversano, Italy, 
14--18 June 2003}

\title{QCD tools for the LHC}
\author{Michelangelo L. Mangano\addressmark{a}}
\address[a]{TH Division, CERN, 1211 Geneva 23, Switzerland}
\begin{document}

\begin{abstract}
This contribution provides a pedagogical introduction to and review of
the current status and ongoing progress in the development of Monte
Carlo tools for the calculation and simulation of high-$Q^2$ processes
in hadronic collisions.
\end{abstract}
\maketitle

\section{Introduction}
In April 2007 the Large Hadron Collider (LHC) will start providing
$pp$ collisions at $\sqrt{S}=14$~\tev. The main goal~\cite{Mangano:jw}
of this new enterprise is the exploration of a yet uncharted energy
domain, and QCD will provide an essential tool for the analysis and
the decoding of the immense set of data that will be collected by the
three detectors currently being built~\cite{TDRS}.  The production
rates for most SM particles and processes are mindboggling. 
Huge statistics of final
states nowadays totally unaccessible at the current accelerators will
be available, and will allow measurements of unprecedented accuracy
and depth~\cite{Altarelli:2000ye}-\cite{Gianotti:2002xx}.  
Table~\ref{tab:rates} gives few examples
of cross-sections for some of the most relevant processes. 
\begin{center}
\vspace*{-1cm}
\begin{table}[ph]
\caption{Benchmark cross-sections for few SM processes at the LHC,
  $L=10^{33}\mathrm{cm}^{-2}\mathrm{s}^{-1}$ . In
  the case of jets and photons, we assume a $\vert \eta \vert <1$
  rapidity cut.}
\label{tab:rates}
{\footnotesize
\begin{center}
\begin{tabular}{|c|r|}
\hline
{} &{} \\[-1.5ex]
Process & $\sigma$ (nb) $\equiv$ evts/s   \\
\hline
 Jets, $E_T>0.1$ (2) TeV & $10^{3}$ ($10^{-4}$)  \\[1ex]
$W^{\pm}\to e\nu_{e}$ & $20$ \\[1ex]
$Z\to e^+e^-$ & $2$ \\[1ex]
Photons ($E_T>60$GeV) & $20$  \\[1ex]
$c \bar c$, $b \bar b$   & $8\times 10^6$, $5\times 10^5$   \\[1ex]
$t \bar t$  & $0.8$ \\[1ex]
\hline
\end{tabular}
\end{center}
}
\vspace*{-13pt}
\end{table}
\end{center}
 In addition to direct manifestations of new phenomena, these
measurements could ultimately lead to indirect evidence for physics
beyond the SM itself. The huge lever arm in energy available through
the measurement of high transverse-energy ($\et$)
jets~\cite{Catani:2000jh} will allow to probe the smallest distance
scales ever accessed. One year of high-luminosity running will give
tens of events with jets with $\et>3$~TeV. Compared to the Tevatron,
where jets up to 600~GeV will be observed, this is a factor of 5
increase in the scale at which the quark form factor can be explored.

The immense samples of EW gauge bosons will enable high-precision
measurements of the $W$ mass ($\pm 15$~MeV) and of the gauge bosons
selfcouplings~\cite{Haywood:1999qg}.  The study of Drell-Yan final
states will be sensitive to several possible new phenomena. For
example, possible contact interactions mixing light quarks with
leptons will be probed up to scales of the order of 25-30~TeV, well in
the region where new strongly interacting phenomena related to the
EWSB may take place. New $U(1)$ gauge bosons with $Z$-like couplings
will be observed up to masses of the order of 4-5~TeV.  The large
statistics will allow measurements of the $W$ and $Z$ total cross
sections with accuracies significantly better than $1\%$, providing
luminosity monitors which are only limited by the precision of the
theoretical predictions for these rates (predictions currently
estimated to be around 5\%, dominated by uncertainties in the
partonic densities of the proton~\cite{Catani:2000jh}).

Top quarks will be produced in great abundance, at a rate of
approximately 1 pair/s. Precise
determinations of the top decay properties and the top mass with an
uncertainty of about 1~\gev\ will be possible~\cite{Beneke:2000hk}.
The study of flavour will be enriched by a thorough
$b$-physics programme~\cite{Ball:2000ba}.  The tens of billions of
bottom quarks produced~\cite{Nason:1999ta}, will allow to pin down
with great accuracy and redundancy the CKM matrix elements, to probe
in full detail the parameters of \cpviol\ in the $B_{q}$ systems
($q=u,d,s$), and to study rare decays with branching ratios at the
level of $10^{-9}$.

In all these cases, it is required that the best possible control on
the QCD production and decay mechanisms be available. For this reason,
in the recent years a great effort has been put into the
development of tools enabling the description of the final states
resulting from high-energy $pp$ collisions.  These tools go under the
common name of Monte Carlo (MC) codes, since the state-of-the-art
knowledge about QCD is implemented using numerical MC techniques.  MC
development is a very technical topic. This review is intended to
provide a pedagogic and qualitative introduction to the main issues,
and to some of the most relevant ideas and topics which are driving
the fast development in the field. More accurate discussions, and all
the necessary details, will have to be looked for in the references.

\section{Overview of the available approaches}
The starting point of all QCD analyses of high-\qsq\ processes is
summarized by the factorization theorem, most clearly expressed by the
following relation:
\ba
\frac{d\sigma}{dX} &=& \sum_{\hat X} \sum_{j,k} \; f_j(x_1,Q_i)\,
f_k(x_2,Q_i) \; \nn \\
&&  \frac{d\hat\sigma_{jk}^{\hat{X}}(Q_i,Q_f)}{d\hat X}
 \; F(\hat X
  \to X;\, Q_f) \; ,
\ea
where: $X$ represents a given hadronic final state (FS);  $\hat X$ is an
arbitrary partonic FS; $f_j(x,Q_i)$ is the number
density of partons of type $j$ carrying the momentum fraction $x$ of
the nucleon at a resolution scale (factorization scale) $Q_i$;
$\hat\sigma_{jk}^{\hat{X}}(Q_i,Q_f)$ is the partonic cross-section for
the transition between the initial partonic state $jk$ and the final
partonic state $\hat X$, considered at resolution scales $Q_i$ and
$Q_f$ for initial and final state factorization,
respectively\footnote{An additional dependence on the renormalization
  scale \mur\ is also present, but this is usually associated and identified
  with $Q_i$ and/or $Q_f$.}; $F(\hat X \to X;\, Q_f)$ represents a
transition function from the partonic final state to the given
observable $X$. This may include fragmentation functions,
hadronization effects, as well as the result of experimental cuts or
jet definitions. The sum over final states $\hat X$ can be thought of
as a sum over all possible histories leading to the same observable
configuration. In a similar fashion, the parton density (PDF)
$f_j(x,Q)$ can be thought of as the sum over all initial-state (IS) evolution
histories leading to the initial-state parton $j$ with momentum
fraction $x$. According to the factorization theorem, the possible IS and FS
histories, and their relative probabilities, are independent of the
hard process, and only depend on the flavours of the partons involved
and on the resolution scales. Once an algorithm is developed to
describe IS and FS evolution, it can therefore be applied to the
partons arising from the calculation of an arbitrary hard
process. Depending on the extent to which possible IS and FS histories
affect the value of $X$, three different realizations of the
factorization theorem are used: cross-section ``evaluators'',
parton-level event generators, and shower MC event generators. We
shall now review these in more detail.
\subsection{Cross-section evaluators}
In this case only some component of the FS is singled out for the
measurement, all the rest being ignored (\ie\ integrated over). One
example is the Drell-Yan (DY) process, where we typically look at the
inclusive spectrum (mass, \pt\, rapidity, etc.) of a leptonic FS. In
this approach, no event needs to be generated, it is enough to define
the variables relative to the considered object (the lepton(s), or a
jet). Experimental selection criteria (\eg\ a jet definition or a
detector acceptance) are applied to parton-level quantities. Provided
these are infrared and collinear finite, it does not matter what
$F$ is, as we assume that all histories of the final partonic state
lead to the same observable, and integrate to 1 because of unitarity:
$  \sum_{\hat X} \; F(\hat X, X) =1$. Thanks to the inclusiveness of
the result, it is conceptually ``straightforward'' to include
higher-order corrections, as well as to resum classes of dominant and
subdominant logarithms. Next-to-leading order (NLO) results are
known~\cite{Catani:2000jh,Giele:2002hx} 
for most processes both within and beyond the Standard Model (SM). In
addition, next-to-next-to-leading order (NNLO) cross-sections have
been calculated for the DY-type processes~\cite{Hamberg:1990np}
 and for Higgs production~\cite{Harlander:2002wh}. For a review of
 progress in NNLO calculations for higher-multiplicity final states, 
see~\cite{Glover:2002gz}.

\subsection{Parton-level event generators}
Here parton-level (PL) configurations (\ie\ states with quarks
and gluons) are generated, with probabilities proportional to the
relative perturbative matrix element (ME). The transition function
between a FS parton and the observed object (jet, missing energy,
lepton, etc.) is unity, so there is no need to model the exclusive
realization of the histories associated to the PDF and to $F$, as
they all lead to the same observable. Experimentally, this is
equivalent to assuming a smart jet algorithm (that would associate a
jet to each hard parton) and linear detector response (the energy and
direction of a measured jet will not depend on its inner
structure). The advantage over the cross-section evaluators is that,
with the explicit representation of the kinematics of all hard objects
in the event, more refined detector analyses can be performed,
implementing complicated cuts and correlations which are otherwise
hard to simulate with the inclusive approach. PL event generators are
typically used to describe final states with several hard jets. Due to
the complexity of the ME evaluation for these many-body
configurations~\cite{Mangano:1991by,Caravaglios:1995cd,Draggiotis:1998gr}, 
calculations are normally available only for
leading-order (LO) cross-sections. In this case, several tools~\cite{Berends:1991ax}-\cite{Tsuno:2002ae} have
recently become 
available, covering all of the necessary processes for signal and
background LHC studies, with jet multiplicities all the way up to 4, 5
or 6, depending on the specific process: $(W\to f\bar{f}')
Q\Qbar+$~jets ($Q$ being a heavy quark, and $f=\ell,q$);
$(Z/\gamma^{*}\to f\bar{f}) \, Q\Qbar+$~jets ($f=\ell,\nu$); $(W\to
f\bar{f}') + \mbox{charm} + N$~jets ($f=\ell,q$); $(W\to f\bar{f}') +
$~jets and $(Z/\gamma^{*}\to f\bar{f})+$~jets; $nW+mZ+lH+$~jets;
$Q\Qbar+$~jets; $Q\Qbar Q'\Qbar'+$~jets, with $Q$ and $Q'$ heavy
quarks; $H Q \Qbar$~jets; $N$~jets; $N\gamma+$~jets. Most of these
processes have been evaluated by more than one group, providing
additional robustness to the calculations and their numerical
implementation. NLO PL event generators are also available for several
low-jet-multiplicity final states, as discussed in
Sect.~\ref{sec:nlo}.

\subsection{Shower MC event generators}
Shower MC generators~\cite{Seymour} provide the most complete
 description of the FS.  Their goal is to generate events consisting
 of physical, measurable hadrons, with a correct description of their
 multiplicity, kinematics and flavour composition. These final states
 can therefore be processed through a complete detector simulation,
 providing the closest possible emulation of a real event.  Shower MC
 codes such as \herwig~\cite{Marchesini:1988cf},
 \pythia~\cite{Sjostrand:1994yb}, \isajet~\cite{Paige:1998xm} or
 \ariadne~\cite{Lonnblad:1992tz} have been known and used for several
 years now, and new tools, \sherpa, are becoming
 available~\cite{Gleisberg:2003xi}.
 
After the generation of a given PL configuration (typically using
a LO ME for $2\to 1$ or $2\to 2$ processes), all possible IS and FS
histories (``showers'') are generated, with probabilities defined by the
shower algorithm. By algorithm, we mean a numerical, Markov-like
evolution, which implements within a given approximation scheme the
QCD dynamics. This includes the probabilities for parton radiation
(gluon emission, or $g\to q\qbar$ splitting), an infrared cutoff
scheme, and a hadronization model. The radiation probabilities
are defined as exclusive quantities, unitarized by the
inclusion of Sudakov form factors. This means that the shower
evolution itself does not alter the overall cross-section, as
estimated from the ME evaluation for the initial hard
process. Therefore a shower MC based on LO matrix elements cannot
provide an estimate of the $K$ factor. Radiation probabilities 
implement leading soft and collinear logarithms, plus some subleading
classes of logs.
Quantum-mechanical correlations between different emissions are
negligible in the case of collinear emission, since in this case
gauges can be defined where the interference
between different diagrams is numerically suppressed; 
in the case of soft emission at large angle, the association to a
specific emitter is ambiguous, as the interference between different
emission diagrams is large. Fortunately, soft emission at large angle
is heavily suppressed, as the interference effects between different
diagrams lead to destructive interference. Quantum coherence can then
be implemented in a branching evolution via, for example, 
 an angular ordering prescription~\cite{Marchesini:1983bm}: if
$\theta_{ij}$ is the angle between two colour-connected partons, the
soft-gluon emission probability from the pair $i,j$ is given by the sum of
two independent emission terms, constrained within angular
regions close to the two emitters:
\be P(\theta_i)\Theta(\theta_{ij}-\theta_i)+
P(\theta_j)\Theta(\theta_{ij}-\theta_j)
\ee
where $\theta_i$, $\theta_j$ are the angles between the soft gluon and
the colour connected partons emitting it, and
$P(\theta)$ is a positive definite probability. 

\section{Complementarity of the three approaches}
Each of the above three approaches has its virtues and shortcomings. Until
recently, we had to select the approach that was closest to our needs,
and live with the fact that it had drawbacks. We first outline a comparison
between the merits of the three approaches, and then review the recent
progress in merging them.

\subsection{Final-state structure}
Shower MC's provide the most complete description. The full
information about the event is available, in terms of physical
hadrons. So this is the only tool
that allows realistic detector simulations, crucial when it comes
to measuring with precision some quantity, such as a particle mass or a
coupling strength. In the case of PL MC's only
the information on the hard partons is given; this is good enough for
 studies with naive detector simulations based on pure
geometry and exact energy resolution. This is an excellent tool to
quickly study systematics such as the PDF dependence of an acceptance, 
but cannot be used, for example,
 as a tool to reconstruct the top quark mass from real $W$+4 jet
 events. Cross-section evaluators
typically don't provide much information on the structure of the final
state.

\subsection{Higher-order corrections: real, hard emissions}
In the shower MC's, emissions are treated within the collinear or soft
approximation. In particular, the angular-ordering prescription,
introduced to implement quantum coherence effects for large-angle
gluon radiation, will suppress hard large-angle emissions as well. Since
large-angle {\em hard} emissions are not constrained by angular
ordering, the shower approximation will typically underestimate the
rate of multijet final states.
On the contrary, PL MC's have all diagram interference
terms among multiple hard partons taken into account, and provide the
best tool to describe multijet final states. 
\subsection{Higher-order corrections: virtual effects}
Virtual diagrams are included in the soft approximation in the shower
MC's. They are needed to cancel the infrared (IR) singularities due to
the real emission of soft gluons, and appear
through the Sudakov form factors to enforce the unitarity of the
shower evolution. Virtual effects are exactly included in
 all NLO cross-section
evaluators. In addition to the exact treatment of real emission, NLO
(or NNLO) cross-section evaluators provide the most accurate
determination of inclusive rates. In order to implement them in ME
event generators, however, one must deal with the problem of
negative-weight events. These arise to enforce the
cancellation of the positive-weight infinities present in the soft and
collinear real
emission diagrams, with negative-weight 
infinities associated to the interference of
the virtual and tree-level diagrams.
\subsection{Resummations}
Resummation~\cite{Dasgupta:2003nh,Catani:2000jh,Giele:2002hx}
of leading and subleading logarithms appearing at all
orders of perturbation theory when largely different energy scales
appear in the event kinematics (for example small-\pt\ production of
heavy objects, with $M\gg \pt$, or large-\pt\ production of light
objects, $\pt\gg m$), is possible in the context of cross-section
evaluators. 
The resummation of logarithms corresponds to the
integration over multiple-emission and multi-loop diagrams, and is
therefore, by its nature, an inclusive calculation. As a result, so
far no ME generator includes them, although new
analytical~\cite{Kidonakis:aq,Bonciani:2003nt} and
semi-numerical~\cite{Banfi:2003je} 
approaches have been proposed to provide resummation corrections to
arbitrary multi-parton final states.  
Shower MC's incorporate
resummations via the multiple emissions taking place in the shower,
and the corresponding Sudakov form factors, which account for the
multi-loops. However, due to the unitarity of the evolution, these
resummations cannot affect the overall rate of a process. The
resummation of threshold-like logarithms (such as those entering in
the determination of the $t\bar t$
cross-section~\cite{Bonciani:1998vc}) affect the rates,
and so far are only included in cross-section evaluator codes.
Resummation of small-$x$ effects\cite{Andersson:2002cf}, finally, 
is implemented in some shower
MC's\cite{Jung:2001hx,Kharraziha:1997dn}, 
mostly developed for HERA studies,
 and is not yet a standard component of LHC tools. 

\section{A Monte-Carlo of Everything}
The efforts of the past few years have
improved considerably the flexibility of the above tools, and have
allowed the construction of MC codes which merge the merits of
different approaches. The main lines of development are summarised, in
a very simplified way, in this section.
\subsection{Inclusion of NLO corrections in ME generators}
\label{sec:nlo}
At NLO one needs to account for both real-emission diagrams and
virtual corrections. Virtual corrections to an N-body FS give
still an N-body FS, while real emissions lead to an (N+1)-body FS. 
An NLO event generator must therefore generate both
N-body  and (N+1)-body FS's. The two however cannot be treated
separately, since the contribution of the virtual
part  to the cross-section 
is equal to minus infinity, while the real part has
divergencies for  collinear and soft configurations. The structure of
the cross-section in these regions can be parameterized in an idealized
form as follows:
\be \frac{d\sigma}{dx} = \frac{f(x)}{(1-x)_+} \; , \ee
where by definition of $()_+$:
\be \label{eq:int}
\int_0^1 dx \frac{d\sigma}{dx} = \int_0^1 dx \; 
\frac{f(x)-f(1)}{(1-x)}\ee
The limit $x\to 1$ corresponds to kinematical configurations where the
(N+1)-body FS degenerates (via soft or collinear emission) to an
N-body FS. In the case of collinear emission, for example,
$x=\cos\theta$. For heavy quark ($Q$) pair production,
$x=m^2_{Q+\Qbar}/\hat{s}\to 1$ corresponds to the soft-emission limit.
$f(x)$ is continuous in $x=1$, and therefore any integral
over an arbitary range of $x$, including possibily $x=1$, is
finite. This is the essence of the virtual/real cancellation of
infinities. If we formally remove the integration from
eq.~\ref{eq:int}, we can ``define'' a non-singular differential
cross-section as:
\be \label{eq:diff}
 \frac{d\sigma}{dx} = 
\frac{f(x)-f(1)}{(1-x)} \; .
\ee
One can interpret this relation as follows: for any given (N+1)-body 
kinematical configuration $C(x)$ ($x\ne 1$), whose
weight is given by $f(x)/1-x$, we can associate an N-body virtual
configuration $C(1)$ with weight $-f(1)/1-x$, whose kinematics is obtained by
the $x\to 1$ limit of $C(x)$. These virtual configurations are
typically called ``counter-events''. One can construct an
event generator by adding to each (N+1)-body final state its
corresponding N-body counter-event. Both events will have a finite weight,
although the counter-event's is negative. Since their kinematics
is different, event and counter-event will typically populate
different bins of our histograms. For bin sizes sufficiently large
(namely for observables sufficiently inclusive), summing events over
the full phase-space will nevertheless lead to positive rates in each bin. 
This is
not guaranteed to happen when the bins are too small. For example, let
us consider the integral of the cross-section in a bin covering the range
$1-\epsilon < x < 1$, with $\epsilon$ small:
\be
\int_{1-\epsilon}^1 d\sigma =  
\int_{1-\epsilon}^1 dx\frac{f(x)}{(1-x)}-\int_0^1 dx\frac{f(1)}{(1-x)}
\; .
\ee
The first contribution is from the $N+1$-body FS's, the second from
the counter-events which all accumulate at $x=1$. Simple algebra leads
to the following result:
\be
\int_{1-\epsilon}^1 d\sigma =  
C +f(1)\log\epsilon \; ,
\ee
where $C$ is a finite constant when $\epsilon\to 0$. Since $f(1)$ is
positive, when the bin-size $\epsilon$ is small enough the integral
becomes negative. This is an indication that radiative corrections in
this small corner of phase-space are large. 
In other words, if we try to push the calculation in a
region of low-inclusivity, probing the final-state structure with very
fine resolution, the fixed-order perturbative approximation breaks
down. Higher-order corrections become very large, and have to be
included to restore the positivity of the cross-section in that
bin. The smaller the bin, the larger the number of orders required.

In concrete applications life is complicated by the identification of
the functions $f(x)$, the interplay of soft and colliner
singularities, and the description of the phase-space in terms of
suitable variables; however the above simplified description captures
the main features of the technique. It has been used, in various
different implementations, for the development of NLO ME event
generators covering several of the interesting LHC processes. 
Examples include:
2-jets~\cite{Giele:dj} and 3-jets~\cite{Nagy:2003tz}, heavy
quarks~\cite{Mangano:jk}, and vector boson~\cite{Giele:1993pk}
production. 
For a
detailed list of available tools, see~\cite{dobbs2004}
The extension of these techniques to NNLO calculations has yet to be
formulated. Work is in progress\cite{Glover:2002gz}, but the
difficulty is immense, and it will still be some time before we can
implement the already known NNLO matrix elements for DY and Higgs
production into an event generator.
\subsection{NLO corrections in shower MC's}
The necessity to include NLO corrections in ME generators is
twofold. On one side, only shower MC's provide a representation of the
final state complete enough to allow realistic detector
simulations. Inclusion of the NLO matrix elements for the hard
process, which will provide cross-sections with full NLO accuracy, is
a natural improvement of these essential tools.  On the other hand, as
mentioned in the previous subsection, the inclusion of NLO effects in
fixed-order ME MC's leads to distributions which are not positive
definite, thus calling for a tool where these large (and possibly
negative) logarithmic effects which arise at any fixed order in some
corners of phase-space can be properly resummed. This goal can be
achieved via the inclusion of the NLO ME's in the shower MC. A priori
one may expect this task to be ill defined, as shower MC's already
incorporate part of the NLO effects: they have real emissions, as well
as virtual effects included in the Sudakov form factors. The naive
introduction of NLO ME's would then lead to double counting. This is
what kept people skeptical for many years about the viability of a NLO
shower MC.  Brilliant work by Frixione and
Webber~\cite{Frixione:2002ik} 
recently showed how
this merging can be done very effectively in what they called 
a \MCNLO.  One starts by
identifying the analytic form of the approximation used by the shower
MC to describe real emission and the leading-order virtual correction
contained in the Sudakov form factor. One can then subtract these
expressions from the NLO matrix elements; since the shower
approximation has the correct residue for all singular contributions,
the subtracted NLO matrix elements are finite. In the simple
formalism used in the previous section, one can represent the
subtraction from the real emission term as follows:
\be 
\frac{d\sigma}{dx} = \frac{f(x)-f_{MC}(x)}{(1-x)} \; , 
\ee 
where $f_{MC}(x)$ is the approximate MC expression for the real
emission matrix element, with the condition that $f_{MC}(1)=f(1)$.
In this way the $x\to 1$ singularity is not removed by the merging with the
virtual correction, but by letting the shower algorithm handle it and
absorb it into the Sudakov form factor. As for the virtual part, the
singular contribution is all contained in the shower approximation,
and what is left for the NLO correction to describe is just a finite
term with the N-body, Born-like, kinematics.  One is still left
with positive and negative weight events, since the difference between
the exact non-singular terms from the full NLO calculation and those
used in the shower can have either sign. However, since
the residual positive and negative weights are bounded, one can define
an unweighting procedure whereby positive-weight events are unweighted
against the maximum positive weight, and negative-weight events are
unweighted against the minimum negative weight. This procedure has
been successfully implemented in \MCNLO\
codes~\cite{Frixione:2002ik,Frixione:2003ei} describing heavy-quark
pair, Higgs, DY 
and gauge boson pair production. 
There is no obstacle
to extending it to all remaining processes known at NLO. Other
approaches have also been developed into alternaitve codes for
DY~\cite{Grace:2003npb}  and
vector boson pair~\cite{Dobbs:2001gb} production.

\begin{figure}[h]
\includegraphics[width=\hsize,clip=]{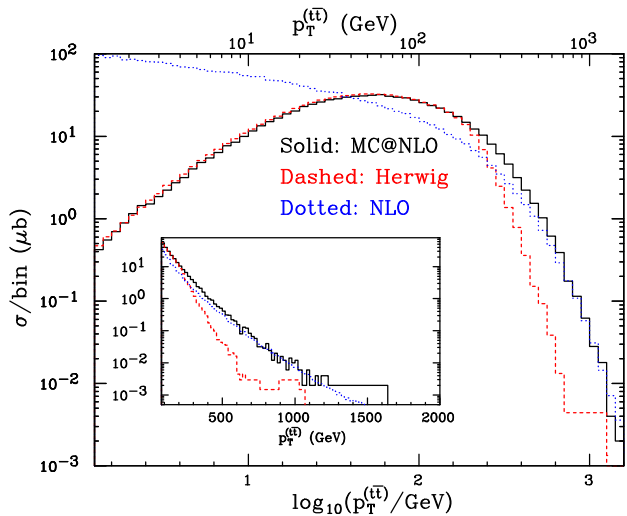}
\caption{Transverse momentum distribution of top quark pairs using
  three different approaches: the LO \herwig\ MC, the parton-level NLO
  MC, and the merging of the two into \MCNLO. Figure
  from~\cite{Frixione:2003ei}. }
\label{fig:hwnlo}
\end{figure}

The inclusion of NLO corrections in the shower MC guarantees that
total cross-sections generated by the MC reproduce those of the NLO ME
calculation, thereby properly including the $K$ factors and
reducing the systematic uncertainties induced by renormalization and
factorization scale variations. At the same time, however, the
presence of the higher-order corrections generated by the shower will
improve the description of the NLO distributions, leading to
departures from the parton-level NLO result. This is shown, for
example, in fig.~\ref{fig:hwnlo}, which shows the \pt\ spectrum of a
$t\bar{t}$ pair resulting from the pure NLO calculation, from the LO
shower, and from the \MCNLO\ improvement. At large \pt, a region
dominated by the NLO effects, \MCNLO\ faithfully reproduces the hard,
large-angle emission distribution given by the NLO matrix elements. At
small \pt, a region dominated by multiple radiation and higher-order
effects, the \MCNLO\ departs significantly from the NLO result, while
properly incorporating the Sudakov resummation effects only available
via the IS shower evolution. 

\subsection{Merging multijet ME generators and shower evolution}
The inclusion of NLO ME's in the shower MC's guarantees the correct
description of the emission of one extra hard parton (ultimately a
jet) from the Born, LO process. In the case of NLO corrections to the
dijet final states, for example, this means that all topologies with
up to three jets will be accurately described. To go beyond this in a
NLO framework, however, requires the knowledge of NLO ME's which are
not available today, and won't be for a longtime to come in the case
of the highest multijet FS's which are of interest for several LHC
studies. As mentioned earlier, the description of multijets obtained
from the shower evolution is inaccurate, since hard radiation at large
angle is suppressed by the angular ordering prescription. One
therefore needs an approach in which multi-parton events generated
using the exact LO ME generator can be consistently evolved into
multi-jet FS's via a shower MC. 
 As in the case of the inclusion of NLO corrections, the
main problem to be dealt with is that of double counting. I shall
describe this problem by discussing a specific example~\cite{mlmtalk}: 
consider
inclusive production of 3-jet events, with jets defined by a cone  of
size $R_j$ in
$\eta-\phi$ space and transverse energy larger than
\def\etmin{\mbox{$E_T^{min}$}} \etmin. We can generate these events by
generating parton-level configurations with partons separated by
$\Delta R>R_j$, and let the shower evolve them into jets. By and
large, there will be a one-to-one correspondence between the generated hard
partons and the jets, and the angular distributions of the three jets
will include correctly all interference effects among the various
diagrams. There will be configurations, however, where the
correspondence is not guaranteed. Take for example events where two
jets have \et\ larger than the third, and the third itself is well
above the \etmin\ threshold: $E_{T1}\sim E_{T2} > E_{T3}\gg
\etmin$. These events can be generated in two independent ways. On one
side we can start from three partons with the kinematics of the three
jets. After evolution, the partons will generate the desired jets. On
the other, we can start from configurations where the two leading
partons have energies of the order of $E_{T1}$ and $E_{T2}$, but the
third can be as soft as \etmin. These events will typically not have a
third jet with $E_T\sim E_{T3}$, since the generated parton will be
softer. However, hard radiation by one of the two leading jets may
lead, with probability \as, to the generation of an extra jet with the
desired energy. In other words, the shower evolution in these cases
could produce jets with transverse energy larger than that of partons
already present in the hard event. While the probability of this
happening is parametrically of order \as\ relative to the LO process,
and could therefore be considered a higher-order correction compatible
with the LO approximation of the ME calculation, 
 configurations with two hard partons and a much softer one are
enhanced by large logarithms. This is the result of the large
phase-space available for the emission of a soft parton in a hard
event. One can estimate that this enhancement is of order
$\log(E_{T3}/\etmin)$. As a result the overall probability that the
third jet be emitted from the shower becomes a number of order $\as
\log(E_{T3}/\etmin)$. On one side this could be numerically of order
1, therefore turning an NLO effect into a LO
one.  On the other, this dependence on \etmin\ of the
cross-section for 3-jet events well above the
\etmin\ threshold is a totally unphysical result!  This
paradox is caused by the double counting of equivalent configurations
which this approach gives rise to. To solve the paradox, and remove
the \etmin\ dependence, one has to carefully check that a given
phase-space configuration is generated only once: either directly by
the parton-level event, or by the shower evolution.

Some approaches to this problem, aiming at different levels of
 accuracy, have been introduced recently . The
first~\cite{Seymour:1995df,Lonnblad:1992tz} 
is generically known as ``matrix-element
correction'' technique
(MEC), as it corrects the approximate ME for the emission of the
 hardest gluon in a given process by using the exact LO ME.
The second is known as CKKW~\cite{Catani:2001cc};
its goal is to implement multi-jet ME
 corrections at the leading (LL), or next-to-leading (NLL) 
logarithmic level. In the MEC, one starts by identifying
analitically the phase-space region $\Omega$ covered by the shower
algorithm. In the case of $e^+e^- \to q\qbar g$, for example, this is
given by a subset of the full phase-space domain $\Delta$ defined by
$\Delta=[1\le x_1+x_2 \le 2] \cap [x_i\le 1] $, where
$x_i=2E_i/\sqrt{S}$. $\Omega$ contains the singular regions
corresponding to soft ($x_i=0$) and collinear ($x_i=1$) gluon
emission.  The integral $\sigma_{\Omega^0}$ of the 3-body cross-section
over the complement of $\Omega$, $\Omega^0=\Delta-\Omega$, is
therefore free of singularities and finite. The integral over
$\Omega$, $\sigma_{\Omega}$, is also finite, once the virtual
corrections at the edge of phase-space are included.  The MEC MC
generation then works by deciding on an event-by-event basis whether
to generate the event in $\Omega^0$ or in $\Omega$, based on the
relative value of the respective cross-sections. If the event falls in
$\Omega$ one generates a LO $e^+e^- \to q\qbar$ event. If it falls in
$\Omega^0$, one generates a $e^+e^- \to q\qbar g$ event. In both
cases, the events are then evolved through the shower. Small
adjustments should then be made in the first case to ensure a proper
continuity across the boundary between the two domains. This technique
has been applied also to DY production~\cite{Corcella:2000gs}, and to
 top decays~\cite{Corcella:1998rs}.  Its
extension to more complicated processes, however, is made particularly
difficult by the need to provide an analytic description of
$\Omega$. When there are more than 3 coloured partons in the process
(as e.g. in dijet or heavy quark pair production at the LHC), this
becomes very hard and impractical. The CKKW approach circumvents this
problem by limiting its precision goal to a LL
accuracy (NLL for $e^+e^- \to$ multijets). 
In this approach the double counting is removed not by
exactly separating a priori the domains $\Omega$ and $\Omega^0$, but
by a probabilistic rejection procedure applied to events falling in
the overlap, so as to ensure that a given phase-space configuration is
only counted once. One starts by generating samples of multi-parton
events of different multiplicities, using the exact LO ME. The
generation is carried out in a phase-space domain defined by a
Durham-like jet algorithm suitable for hadronic
collisions~\cite{Catani:1993hr}; a
resolution variable for two partons is defined by: 
\be 
k_{ij}=\mathrm{min}(E_{T,i},E_{T,j}) R_{ij} \; , 
\ee 
if both $i$ and $j$ are in the FS, or by $k_{iJ}=E_{T,i}$ if $J$ is in
the IS. $R_{ij}$ is a measure of separation in the transvserse plane,
for example the standard $\sqrt{\Delta\phi^2+\Delta\eta^2}$ measure.  An
$N$-parton FS is then classified as an $N$-jet event if
$k_{ab}>k_{cut}$ for all possible parton pairings. $k_{cut}$ is a
resolution threshold, introduced as a parameter necessary to separate
 the generation of the events in samples of different jet
 multiplicity; the final cross-setion, however, should be indepepdent
 of its specific values.  Events are extracted from the different $N$-jet
samples with probability proportional to the sample cross-section. The
jet algorithm can then be used to define a tree structure for the
event. The two partons $i,j$ with the smallest $k_{ij}$ are clustered
into a single virtual parton $\ell$, provided the parton types and
flavours of $i$ and $j$ can be merged. The procedure is repeated after
removing $i,j$ from and adding $\ell$ to the list of partons. The
clustering continues until one gets a $2\to 1$ or $2\to 2$
process. The resulting tree is then interpreted as a shower
configuration, but with a weight given by the exact ME. To be used as
an exclusive FS, and to allow the successive evolution via the shower,
the event weight needs to be corrected with the inclusion of Sudakov
form factors for each vertex, and by a rescaling of the values of \as\
reflecting the choice of scale for \as\ made by the shower.  This
reweigthing factor, which can be constructed so as to always be
smaller than 1, is then used as a probability to keep or reject the
event. The events kept are then showered, and shower splittings at a
scale larger than $k_{cut}$ are vetoed, to avoid the duplication of
configurations which will otherwise be present in ME samples
correpsonding to higher $N$.  In the case of $e^+e^-$
 collisions~\cite{Catani:2001cc,Lonnblad:2001iq}  one
can prove that this algorithm correctly reproduces the weight of an
event to NLL accuracy, and to rates
independent of $k_{cut}$. In hadron collisions~\cite{Krauss:2002up}
 such a proof is still missing, but this framework provides a very good
starting point for further developments. The results of the first
studies in the case of $W$+jets production can be found in~\cite{kraussetal}.

An additional improvement in the description of multijet final states
may soon come from the development of new shower
algorithms,  being developed in the context of the
new generation of C++ codes, such as \herwig++~\cite{Gieseke:2003hm}
and \sherpa~\cite{Gleisberg:2003xi}.  Preliminary results for $e^+e^-$
collisions, comparisons with LEP/SLC data and examples of the improved
description of multijet final states, can be found
in~\cite{Gieseke:2003hm}.  Similar work is under way for the C++
version of \pythia~\cite{Bertini:2000uh}.


\section{Conclusions}
I hope to have suceeded in giving a flavour of the big progress and
flourishing of new ideas which is currently taking place in the field
of MC development for hadronic collisions. Tools only dreamt of few
years ago are now available, encouraging and justifying greater
ambitions. Progress is also taking place in other directions which
I have not had time to cover, such as the progress in the evaluation
of NNLO matrix elements~\cite{Glover:2002gz}, in the description of
power corrections~\cite{Banfi:2001aq},
underlying event and multiple interactions~\cite{Skands:2003yn},
hadronization~\cite{Winter:2003tt},
as well as on the uniformization of input/output
formats for the merging of PL and shower generators~\cite{Boos:2001cv}
The validation of MC tools against Tevatron data is also an area of
active and successfull research~\cite{Field:2002vt,Field:2002da}.
Finally, great progress has been achieved in the
determination of parton densities and their
sistematic uncertanties~\cite{Giele:2001mr}, 
an essential input for all event
simulations. Hoping that the progress will continue, and that all of
these tools will receive proper validation using the results available
from the Tevatron, I look forward to their testing at the LHC.

\section*{Acknowledgments}
I am grateful to the organizers, and to
Prof. Nardulli in particular, for the kind invitation to attend this
very interesting and successfull meeting.

\end{document}